\input harvmac
\noblackbox
 
\font\ticp=cmcsc10
 
\def\Title#1#2{\rightline{#1}\ifx\answ\bigans\nopagenumbers\pageno0\vskip1in
\else\pageno1\vskip.8in\fi \centerline{\titlefont #2}\vskip .5in}

\font\ticp=cmcsc10
\font\ttsmall=cmtt10 at 8pt

\input epsf
\ifx\epsfbox\UnDeFiNeD\message{(NO epsf.tex, FIGURES WILL BE
IGNORED)}
\def\figin#1{\vskip2in}
\else\message{(FIGURES WILL BE INCLUDED)}\def\figin#1{#1}\fi
\def\ifig#1#2#3{\xdef#1{fig.~\the\figno}
\goodbreak\topinsert\figin{\centerline{#3}}%
\smallskip\centerline{\vbox{\baselineskip12pt
\advance\hsize by -1truein\noindent{\bf Fig.~\the\figno:} #2}}
\bigskip\endinsert\global\advance\figno by1}

%
%
\def\[{\left [}
\def\]{\right ]}
\def\({\left (}
\def\){\right )}
\def\p{\partial}
\def\R{{\bf R}}
\def\S{{\bf S}}
\def\l{\ell}
\def\g{\gamma}

\def\e{\varepsilon}
\def\Om{\Omega}
\def\a{\alpha}
\def\b{\beta}
\def\lam{\lambda}

\def\CP{{\cal P}}
\def\CV{{\cal V}}

\def\ud{\dot{u}}
\def\vd{\dot{v}}

\def\xid{\dot{x}^i}

\def\dua{\( {\p \over \p u} \)^a}
\def\dva{\( {\p \over \p v} \)^a}

\def\dya{\( {\p \over \p y} \)^a}

\def\itm{\noindent $\bullet$ \ }
\def\pinf{p_{\infty}}

\def\vinf{v_{\infty}}
\def\xinf{x^i_{\infty}}
\def\Du{\Delta u}
\def\Dv{\Delta v}
\def\Dxi{\Delta x^i}
\def\ie{{\it i.e.\ }}
\def\eg{{\it e.g.\ }}


\lref\PWBS{
E.~G.~Gimon, A.~Hashimoto, V.~E.~Hubeny, O.~Lunin and M.~Rangamani,
{\it Black strings in asymptotically plane wave geometries},
JHEP {\bf 0308}, 035 (2003)
arXiv:hep-th/0306131.}

\lref\NullSoln{
V.~E.~Hubeny and M.~Rangamani,
{\it Generating asymptotically plane wave spacetimes},
JHEP {\bf 0301}, 031 (2003)
[arXiv:hep-th/0211206].}

\lref\NoHor{
V.~E.~Hubeny and M.~Rangamani,
{\it No horizons in pp-waves},
JHEP {\bf 0211}, 021 (2002)
[arXiv:hep-th/0210234].}

\lref\maro{
D.~Marolf and S.~F.~Ross,
{\it Plane waves: To infinity and beyond!},
[arXiv:hep-th/0208197].}

\lref\penrose{
R.~Penrose,
``Any spacetime has a planewave as a limit,'' in
{\it Differential geometry and relativity}, pp 271-275,
Reidel, Dordrecht, 1976.}

\lref\bmn{
D.~Berenstein, J.~M.~Maldacena and H.~Nastase,
{\it Strings in flat space and pp waves from N = 4 super Yang Mills},
JHEP {\bf 0204}, 013 (2002)
[arXiv:hep-th/0202021].}

\lref\bfhp{
M.~Blau, J.~Figueroa-O'Farrill, C.~Hull and G.~Papadopoulos,
{\it A new maximally supersymmetric background of IIB superstring theory},
JHEP {\bf 0201}, 047 (2002)
[arXiv:hep-th/0110242].}

\lref\bena{
D.~Berenstein and H.~Nastase,
{\it On lightcone string field theory from super Yang-Mills and holography},
[arXiv:hep-th/0205048].}

\lref\magoo{
O. Aharony, S.S. Gubser, J. Maldacena, H. Ooguri, Y. Oz,
{\it Large N Field Theories, String Theory and Gravity},
Phys. Rept. {\bf 323} (2000) 183, [arXiv:hep-th/9905111].}

\lref\gava{
D.~Garfinkle and T.~Vachaspati,
{\it Cosmic String Traveling Waves},
Phys.\ Rev.\ D {\bf 42}, 1960 (1990).}

\lref\garfinkle{
D.~Garfinkle,
{\it Black String Traveling Waves},
Phys.\ Rev.\ D {\bf 46}, 4286 (1992)
[arXiv:gr-qc/9209002].}

\lref\coley{
A.~A.~Coley,
{\it A class of exact classical solutions to string theory},
Phys.\ Rev.\ Lett.\  {\bf 89}, 281601 (2002)
[arXiv:hep-th/0211062].}

\lref\pravda{
V.~Pravda, A.~Pravdova, A.~Coley and R.~Milson,
{\it All spacetimes with vanishing curvature invariants},
Class.\ Quant.\ Grav.\  {\bf 19}, 6213 (2002)
[arXiv:gr-qc/0209024].}

\lref\coleyetal{
A.~Coley, R.~Milson, N.~Pelavas, V.~Pravda, A.~Pravdova and R.~Zalaletdinov,
{\it Generalized pp-wave spacetimes in higher dimensions},
Phys.\ Rev.\ D {\bf 67}, 104020 (2003)
[arXiv:gr-qc/0212063].}

\lref\bcr{
D.~Brecher, A.~Chamblin and H.~S.~Reall,
{\it AdS/CFT in the infinite momentum frame},
Nucl.\ Phys.\ B {\bf 607}, 155 (2001)
[arXiv:hep-th/0012076].}

\lref\gy{
G.~T.~Horowitz and H.~S.~Yang,
{\it Black strings and classical hair},
Phys.\ Rev.\ D {\bf 55}, 7618 (1997)
[arXiv:hep-th/9701077].}

\lref\alga{
M.~Alishahiha and O.~J.~Ganor,
{\it Twisted backgrounds, pp-waves and nonlocal field theories},
JHEP {\bf 0303}, 006 (2003)
[arXiv:hep-th/0301080].}

\lref\giha{
E.~G.~Gimon and A.~Hashimoto,
{\it Black holes in Goedel universes and pp-waves},
Phys.\ Rev.\ Lett.\  {\bf 91}, 021601 (2003)
[arXiv:hep-th/0304181].}

\lref\baro{
V.~Balasubramanian and S.~F.~Ross,
{\it Holographic particle detection},
Phys.\ Rev.\ D {\bf 61}, 044007 (2000)
[arXiv:hep-th/9906226].}

\lref\lmr{
J.~Louko, D.~Marolf and S.~F.~Ross,
{\it On geodesic propagators and black hole holography},
Phys.\ Rev.\ D {\bf 62}, 044041 (2000)
[arXiv:hep-th/0002111].}

\lref\maldacena{
J.~M.~Maldacena,
{\it Eternal black holes in Anti-de-Sitter},
JHEP {\bf 0304}, 021 (2003)
[arXiv:hep-th/0106112].}

\lref\kos{
P.~Kraus, H.~Ooguri and S.~Shenker,
{\it Inside the horizon with AdS/CFT},
Phys.\ Rev.\ D {\bf 67}, 124022 (2003)
[arXiv:hep-th/0212277].}

\lref\lero{
T.~S.~Levi and S.~F.~Ross,
{\it Holography beyond the horizon and cosmic censorship},
Phys.\ Rev.\ D {\bf 68}, 044005 (2003)
[arXiv:hep-th/0304150].}

\lref\fhks{
L.~Fidkowski, V.~Hubeny, M.~Kleban and S.~Shenker,
{\it The black hole singularity in AdS/CFT},
arXiv:hep-th/0306170.}

\lref\causal{
V.~E.~Hubeny and M.~Rangamani,
{\it Causal structures of pp-waves},
JHEP {\bf 0212}, 043 (2002)
[arXiv:hep-th/0211195].}

\lref\penrose{
R.~Penrose,
{\it Any spacetime has a planewave as a limit}, in
{\it Differential geometry and relativity}, pp 271-275,
Reidel, Dordrecht, 1976.}

\lref\bfhp{
M.~Blau, J.~Figueroa-O'Farrill, C.~Hull and G.~Papadopoulos,
{\it A new maximally supersymmetric background of IIB superstring theory},
JHEP {\bf 0201}, 047 (2002)
[arXiv:hep-th/0110242].}

\lref\lpv{
J.~T.~Liu, L.~A.~Pando Zayas and D.~Vaman,
{\it On horizons and plane waves},
[arXiv:hep-th/0301187].
}

\lref\ops{
N.~Ohta, K.~L.~Panigrahi and Sanjay,
{\it Intersecting branes in pp-wave spacetime},
[arXiv:hep-th/0306186].
}

\lref\bdgo{
D.~Brecher, U.~H.~Danielsson, J.~P.~Gregory and M.~E.~Olsson,
{\it Rotating black holes in a Goedel universe},
JHEP {\bf 0311}, 033 (2003)
[arXiv:hep-th/0309058].
}

\lref\kmr{
N.~Kaloper, R.~C.~Myers and H.~Roussel,
{\it Wavy strings: Black or bright?},
Phys.\ Rev.\ D {\bf 55}, 7625 (1997)
[arXiv:hep-th/9612248].
}

%
\baselineskip 16pt
\Title{\vbox{\baselineskip12pt
\line{\hfil SU-ITP-03/30}
\line{\hfil UCB-PTH-03/31}
\line{\hfil LBNL-54019}
\line{\hfil \tt hep-th/0311053} 
}}
{\vbox{
{\centerline{Horizons and plane waves: A review}
}}}
\centerline{\ticp Veronika E. Hubeny$^a$
 and Mukund Rangamani$^{b,c}$ \footnote{}{\ttsmall
 veronika@itp.stanford.edu, mukund@socrates.berkeley.edu}}
\bigskip
\centerline {\it $^a$
Department of Physics, Stanford University, Stanford, CA 94305, USA} 
\centerline{\it $^b$ Department of Physics, University of California,
Berkeley, CA 94720, USA} 
\centerline{\it $^c$ Theoretical Physics Group, LBNL, Berkeley, CA 94720, USA}

\bigskip
\centerline{\bf Abstract}
\bigskip

\noindent

We review the attempts to construct black hole/string solutions in 
asymptotically plane wave spacetimes. First, we demonstrate 
that  geometries admitting a covariantly constant null Killing 
vector cannot admit event horizons, which implies that pp-waves
can't describe black holes.  However, relaxing the 
symmetry requirements allows us to generate solutions which do possess 
regular event horizons while retaining the requisite asymptotic
properties. In particular, we present two solution generating techniques 
and use them to construct asymptotically plane wave black string/brane 
geometries.

\Date{November, 2003}
%
\newsec{Introduction}
It is an incontrovertible fact that black holes
provide an important window into quantum gravity.
Knowledge of exact black hole solutions and their properties has 
greatly enhanced our understanding of gravitational physics
and provided us with valuable intuition. Many of the 
puzzles and paradoxes associated with the unification of 
general relativity and quantum mechanics have their origins 
in the fascinating behaviour of black hole event horizons and 
hence, while we are aware of a rich set of solutions,  
it is always desirable to extend this list.

Black hole solutions in asymptotically ``nice'' (\eg\ maximally
symmetric) spacetimes, such as Minkowski, de Sitter, and Anti de Sitter,
have been known for decades, and have proved very fruitful in furthering
our understanding. A good example is the Schwarzschild-AdS black
hole, whose intricacies one may delve into using the 
well-known AdS/CFT correspondence \magoo.  
Indeed, much recent attention has been devoted to probing 
behind-the-horizon physics, and especially the singularity, via the CFT 
\refs{\baro,\lmr,\maldacena,\kos,\lero,\fhks}.
Unfortunately, all of these efforts have been hampered by the
fact that the AdS/CFT is a strong-weak coupling duality; in the regime
where supergravity is valid, the gauge theory is strongly coupled, 
and therefore computationally intractable.  

Plane waves, or more generally pp-waves, are another class of spacetimes
with rather special properties.  Since they admit a covariantly
constant null Killing vector, all their curvature invariants vanish
identically, rendering these spacetimes exact solutions to classical
string theory (with vanishing $\a'$ corrections).
Nevertheless, they are causally quite nontrivial. 
For instance, remarkably enough, 
a large class of these has a one-dimensional null boundary
\refs{\bena,\maro,\causal}.

It would be therefore very intriguing if these solutions could also
admit event horizons, for this would yield exact black hole solutions
in classical string theory with rather nontrivial asymptotics.
Unfortunately, it turns out that this is not possible, as shown
in \NoHor\ and reviewed below: pp-waves cannot admit event horizons.
This however does not mean that there cannot exist black hole solutions
with pp-wave asymptopia.  After all, in all of the maximally symmetric
spacetimes (flat, dS, and AdS), one also has to break some part of the 
symmetry in order to obtain black holes in these spacetimes.
Indeed, below 
we will review two distinct methods for obtaining black string solutions
which are asymptotically plane wave. 

Apart from these solutions representing black holes with
interesting asymptopia, they serve as important steps towards 
extending the recent BMN correspondence \bmn.
The BMN correspondence may be viewed as a particular limit of the AdS/CFT
correspondence, which on the gravity side generates a maximally 
supersymmetric plane wave, while on the gauge theory side limits
to a large charge sector.  This near-BPS regime allows one to 
do perturbative calculations on both sides of the correspondence, 
thus allowing greater computational control.  It would be therefore
very desirable to use this feature to probe the
mysteries of quantum gravity, and as the first step, to study black holes
in this set-up.  

In particular, one may hope that a black hole with the
particular plane wave asymptotics considered by \bmn\ would correspond
to some tractable modification on the gauge theory side, which we could then
use to study such deep issues as singularity resolution, information 
paradox, etc. So far, such hope is founded mostly on analogy with similar
extension of the original AdS/CFT correspondence:  Putting a Schwarzschild
black hole in AdS spacetime corresponds to thermalizing the gauge theory.
Although the BMN regime contains conceptual puzzles of its own, 
studying black holes in this set-up would be an important endeavour.
The requisite solutions have not yet been found; nevertheless, one may 
be optimistic, as the solutions discussed below already come quite close.
In particular, metrically they have the desired asymptotics, though the
matter support is different from that arising in the BMN correspondence.

This work does not contain any essentially new results; rather, it summarizes
the main points of a series of papers \NoHor, \NullSoln, and \PWBS, all 
motivated by the above goal.  
We start in the next section by setting notation and 
establishing the concept of black holes in pp-waves.
We then present a simplified proof that pp-waves cannot represent black
holes\foot{
We use the terminology black ``holes'' loosely, to denote all black objects, 
such as black strings/branes.
} in Section 3.  The following two sections discuss black strings
in {\it asymptotically} plane wave spacetimes: extremal solutions with
vacuum plane wave asymptotics in Section 4, and more robust, many-parameter
family of black string solutions (including Schwarzschild-like one) with
maximally symmetric plane wave asymptotics in Section 5.  
We end with a discussion in Section 6.

\newsec{Definitions}

To set the notation and re-emphasize terminology, we will first 
write down the pp-wave and plane wave metrics, and then discuss
what we mean by a black hole with the corresponding asymptopia.

The pp-wave spacetimes are defined as spacetimes admitting 
a covariantly constant null Killing field.
The metric in $d$ dimensions can be written as
\eqn\pp{
ds^2 = -2 \, du \, dv - F(u, x^i) \, du^2 + dx^i \, dx^i \ ,}
where $F$ is an arbitrary
function of $u$ and the transverse coordinates $x^i$ with
 $i=1,\cdots,d-2$.
The fact that $\dva$ is a covariantly constant null Killing field
ensures that all curvature invariants vanish in these spacetimes.
Similarly, in the context of string theory, all the $\a'$ corrections
vanish.

Plane waves are a special subset of pp-waves, wherein
the function $F$ is further restricted to be quadratic in the 
transverse coordinates, while still maintaining arbitrary
$u$ dependence.  In particular, plane waves can be written as
\eqn\plane{
ds^2 = -2 \, du \, dv - f_{ij}(u) \, x^i x^j \, du^2 + dx^i \, dx^i \ .}
These solutions may be thought of as arising from a Penrose limit \penrose,
which essentially zooms in on any null geodesic of any spacetime 
and reexpands the transverse coordinates.  This induces an extra 
``planar'' symmetry along the wavefronts.
Further restricting $f_{ij}(u)$ can enhance this symmetry 
considerably; of greatest import will be the maximally symmetric plane waves,
for which $f_{ij}(u)= \mu^2 \, \delta_{ij}$, and which have a large isometry 
group.  A particular example of this type of solution is the BFHP plane wave 
in 10-dimensions \bfhp.

We have discussed the metric, but to specify the full solution,
one also needs to specify the matter content.
Since the only nonvanishing component of the Ricci tensor is
$R_{uu} = {1 \over 2 } \nabla^2_T \, F(u,x^i)$, we see that for
vacuum spacetimes, $F(u,x^i)$ must be harmonic in the transverse
coordinates.  For plane waves, this translates into $f_{ij}$ being
traceless, which in turn implies that nontrivial vacuum plane waves cannot
be spherically symmetric; and conversely, the maximally symmetric
plane waves must have some matter support\foot{
Below, we will use the notation $\CV_d$ to denote the vacuum plane 
waves, and $\CP_d$ to denote the maximally symmetric plane waves, 
in $d$ dimensions.}. 

Nevertheless, to ascertain the presence of black holes (\ie to
find whether or not a given spacetime admits an event horizon),
it of course suffices to consider the properties of the metric 
alone, since that is what determines the causal properties of the 
solution.  Accordingly, in the next section, we will consider 
general pp-wave metrics without requiring them to be solutions
of any particular theory.

Before proceeding to prove the absence of black holes
in pp-waves, we first need to specify what we mean by black holes.
A black hole, by definition, is a region inside event horizon; but
an event horizon is well defined only in asymptotically flat 
spacetime, namely as the boundary of the causal past of the future
null infinity ({\it scri }).  
In other words, a given asymptotically flat spacetime
cannot admit an event horizon if the past of {\it scri}
contains the full spacetime (\ie from every point of the spacetime,
there exists a causal curve which reaches {\it scri}).  
In more general spacetimes, lacking universally-defined asymptotics,
we can try to follow the spirit of this definition by replacing 
``{\it scri}'' with ``arbitrarily far'' in spatial directions.  
While this is generally somewhat murky definition, as we explained
in \NoHor, it may be adopted for the pp-waves \pp\ as the following
working definition:

\noindent
{\bf Def:} 
A pp-wave spacetime does not admit an event horizon
iff  from any  point in the spacetime, 
say $(u_0, v_0, x^i_0)$, there exists a future-directed causal curve
to some point $(u_1,v_\infty, x^i_\infty)$, where $u_1>u_0$ is arbitrary, 
while $v_\infty, x^i_\infty \rightarrow \infty$. 

\noindent
The important aspect is that not just $v$, but also at least one of the
transverse coordinates, $x^i$, gets arbitrarily big along a causal 
curve\foot{
Since the fact that $\dva$ is null and Killing implies that it describes
orbits of null geodesics moving in the $v$ direction, the criterion 
that from any point in such a spacetime there is a causal curve attaining
arbitrarily large $v$ is trivially satisfied.
Furthermore, the crucial fact that the coordinate chart used in \pp\ 
covers the full spacetime makes this definition meaningful.}.
We will in fact use a stronger version of this criterion; namely, we 
will require $u_1 = u_0 + \e$, for arbitrarily small $\e > 0$.
This allows us to use the criterion in greater generality, 
in particular even in the cases where the spacetime terminates at
some finite $u$, {\it i.e.}, $F(u,x^i) \to \infty$
as $u \to u_{\infty} < \infty$.

\newsec{No horizons in pp-waves}

We will now use the above definition to demonstrate why pp-waves
cannot admit horizons, and therefore cannot be black holes.
This was motivated for plane waves and proved for general pp-waves
in \NoHor; here we will present a simplified version of the
proof, which applies for all spacetimes \pp\ with $F(u,x^i) \ge 0$,
and refer the reader to \NoHor\ for the more general proof.

The lack of horizons in plane waves is very simple to understand 
intuitively: plane waves have a planar symmetry which precludes
any special position at which the horizon could be\foot{
This of course doesn't mean that there can't be observer-dependent
horizons, such as in Rindler space or de Sitter; however, since 
Rindler horizons exist in any causally well-behaved spacetime, 
here we shall be
concerned only with the black hole ``event'' horizons. 
}.  Alternately,
one can argue that any plane wave is a Penrose limit of some 
spacetime, but event horizons cannot be retained under a Penrose 
limit, as the latter loses the global information about the spacetime.
Hence, we will present a simplified version of the no-horizon proof 
for the more general pp-wave spacetimes; 
the fact that plane waves can't have horizons then follows
immediately as a corollary.

As indicated above, in order to demonstrate the absence of horizons, 
it suffices to show that from {\it any } point 
$p_0=(u_0,v_0, x^i_0)$ of  the spacetime,
there exists a future-directed, causal curve $\g$ (\ie one along which
$\ud > 0$ and 
$-2 \,\ud \, \vd -F(u,x^i) \, \ud^2 + \xid \, \xid \le 0$)
which reaches arbitrarily large values of $r$ and $v$ in arbitrarily small
$\Delta u$. One can construct such a curve $\g$ explicitly, as demonstrated
in \NoHor; however, there is a more powerful and elegant technique
which proceeds by introducing a fiducial metric, with smaller
``light cones'' than those of the physical spacetime we are interested in.
Any curve which is causal in the fiducial metric will then also have
to be causal in the physical one.  The advantage of this observation
is that by selecting appropriate fiducial metric, the construction 
of causal curves with the requisite properties is rendered much easier.

If $F(u,x^i)$ is non-negative everywhere, such a convenient fiducial metric
is simply the flat spacetime, which corresponds to the pp-wave metric
\pp\ with $F(u,x^i) \equiv 0$.
This translates into finding a curve $\g(\lam)$ such that
\eqn\conds{\eqalign{
& \g(0) = (u_0,v_0, x^i_0)   \cr
& \g(1) = (u_0+\e ,\vinf, \xinf)   \cr
{\rm and } \ \ \ \ 
& -2 \, \ud \,  \vd  + \xid \, \xid \le 0}}
In flat spacetime, this is easily achieved by considering
a ``straight line'' between the points $p_0 \equiv \g(0)$ and 
$\pinf \equiv \g(1)$, and the causal relation simply integrates to
\eqn\causal{
-2 \, \Du \, \Dv + \Dxi \, \Dxi \le 0}
The desired causal curve can then be obtained by the following series
of steps:

\itm
Pick any point $p_0=(u_0,v_0, x^i_0)$ in the spacetime.

\itm
Choose arbitrarily large $\xinf$ (or equivalently, arbitrarily large $\Dxi$).

\itm 
Choose $\Du = \e$ arbitrarily small (but positive, so that $\g$ is
future-directed).

\itm
Since \causal\ implies that $\g$ will be causal if
$\Dv \ge {\Dxi \, \Dxi \over 2 \,\e}$, choose arbitrarily large $\vinf$
such that
\eqn\vcond{
\vinf \ge {\Dxi \, \Dxi \over 2 \, \e} + v_0 \ . }

This construction then automatically gives the requisite causal curve
in the physical spacetime \pp\ as well, thereby proving that pp-waves
cannot admit event horizons.   As mentioned earlier, \NoHor\ generalizes
this proof to arbitrary $F(u,x^i)$.

The no-go theorem for horizons in pp-wave metrics is rather disappointing
from the standpoint of finding black holes in string theory described by
exact classical solutions which receive no $\a'$ corrections.  
However, this does not necessarily imply that there are no such solutions,
since the pp-waves are not the most general spacetimes with vanishing
curvature invariants; cf.\ \eg
\refs{\pravda,\coley,\coleyetal}.  It would be interesting to repeat
our analysis for these more general spacetimes.

\newsec{Horizons in spacetimes with null isometry}

Having shown that spacetimes with a covariantly constant null
Killing field (pp-waves) do not admit event horizons, we can now
ask whether relaxing the covariantly constant requirement 
enables us to construct spacetimes with horizons which nevertheless
have a globally null Killing vector.  
Naively, one might expect 
that even null isometry is too restrictive to allow for horizons;
but this is clearly not the case, as there are well-known examples
of extremal asymptotically flat black branes with null isometry.

Although these solutions are asymptotically flat, it is not difficult
to write down asymptotically vacuum plane wave black branes, 
as discussed in \NullSoln\ and as we briefly review below.
The ``trick'' is to use a solution generating technique, originally
developed by \refs{\gava,\garfinkle}, 
called the Garfinkle-Vachaspati (GV) method.
Given a solution 
to Einstein's equations (in general, with some appropriate 
matter content), which admits
a hypersurface-orthogonal null Killing field compatible
with the matter content, one 
can deform the solution to a new one with the same matter
fields. In particular, the curvature invariants of the deformed 
solution are identical to the parent solution. The  
idea is that given a solution with an appropriate set of symmetries,
one can essentially ``linearize'' Einstein's equations, which will then 
allow one to superpose solutions.

A trivial example of using the GV technique is to apply it to flat space
to obtain a vacuum plane wave, or more generally, a vacuum pp-wave.
In particular, starting with the metric \plane\ (with $f=0$ for flat 
space), the GV solution generating technique essentially amounts to 
adding a term $F(u,x^i) \, du^2$, where $F$ is harmonic, to the metric.  
This by definition corresponds to a pp-wave.   
To use the GV construction for the problem at hand, the strategy is to start
with a solution with a regular horizon and add a term which preserves 
the integrity of the horizon, but changes the asymptotics in the required
fashion.  As mentioned above, the (asymptotically flat) extremal black
branes in ten (IIA/IIB) or eleven dimensional supergravity
satisfy the necessary criteria for the GV construction to be 
applicable; but not all have a regular horizon.  Specifically, 
only the D3, M2, and M5 branes have a regular horizon, and of these, 
only the M2-brane solution is capable of producing a spacetime wherein 
the regular event horizon cloaks a singularity; so we will restrict our
attention to this case (the other cases were briefly discussed in \NullSoln).

To deform the M2-brane 
solution in eleven dimensional supergravity so that the asymptotic
behaviour is $\CV_{10} \times \R$, whilst retaining the nature of the  
near-horizon geometry and the singularity intact, we proceed as follows.
We start with the M2-brane solution to 11-dimensional supergravity, given by
\eqn\mtwosoluv{\eqalign{
ds^2 &= H(r)^{-{2 \over 3}} \, \( -2 \, du \, dv+ dx^2 \) + H(r)^{{1\over 3}}
\, \( dr^2 + r^2 \, d\Om_7^2 \) \cr
F_4 &= \({d H(r)^{-1} \over dr }\) \, du \wedge dv \wedge dx \wedge dr.
}}
where we have combined time with one of the spatial directions along
the brane to write the metric in a more convenient form; to wit,
 both $\dva$ and $\dua$ are null Killing vectors and we can also 
verify that they are hypersurface-orthogonal. 
$H(r) = 1 + {Q^6 \over r^6}$ is a harmonic function in the transverse 
eight-dimensional space. The horizon is at $r =0$ in these coordinates. 
To see the location of the singularity, it is best to define a new coordinate
$\zeta = r^2$. The singularity is located at the zero of $H(\zeta)$, 
{\it i.e.}, $ \zeta = - Q^2$.

We can now apply the GV construction to write a new (``deformed'') metric as
\eqn\gvmtwodef{
ds^2 = H(r)^{-{2 \over 3}} \, \( -2 \, du \, dv+ dx^2  - 
\Psi(u,x,r,\Om_7) \, du^2
\) + H(r)^{{1\over 3}}
\, \( dr^2 + r^2 \, d\Om_7^2 \) 
}
where the new term $\Psi$ must be harmonic, $\nabla^2 \, \Psi = 0$.
Writing $\Psi(u,x,r,\Om_7) = \xi_{kL}(u) 
\, e^{i k x} \, \psi_{k \l}(r) \, 
Y_L(\Om_7) $ with arbitrary functions $\xi_{kL}(u)$, 
and $Y_L(\Om_7)$ denoting the spherical harmonics with $L$ being a 
label for the set of angular momenta on the seven sphere with principal 
angular momentum $\l$, we find the radial equation
\eqn\psimtworad{
 {d^2 \psi_{k \l}(r) \over dr^2 } +
{7 \over r } \, {d\psi_{k \l}(r) \over dr} - \({\l(\l+6) \over r^2} + 
k^2 \, H(r) \)\, \psi_{k \l}(r)   =0
}

For the case of $ k =0$, {\it i.e.}, requiring $\( \partial \over \partial x
\)^a$ to be a Killing vector in the deformed geometry, the problem 
reduces to solving the Laplace equation in eight dimensional flat space.
So we can pick the $\l =2 $ mode on the seven sphere to obtain a 
solution which is asymptotically plane wave.
For example, parameterizing the seven sphere by the coordinates 
such that $\theta$ corresponds to the azimuthal angle, we can choose
$\Psi(u,x,r,\Om_7) = r^2 \, \(8 \cos^2 \theta -1 \)$. Then
\eqn\mtwodefsol{\eqalign{
ds^2 &= H(r)^{-{2 \over 3}} \, \[ -2 \, du \, dv+ dx^2  - 
r^2  \( 8 \, \cos^2\theta -1 \) \,du^2
\] \cr 
& \;\;\;\;\;\;\;\; + H(r)^{{1\over 3}}
\, \[ dr^2 + r^2 \( d\theta^2 + \sin^2\theta \, d\Om_6^2 \) \] \cr
F_4 &= \({d H(r)^{-1} \over dr }\) \, du \wedge dv \wedge dx \wedge dr,
}}
is a solution to eleven dimensional supergravity, with the same 
curvature invariants as the M2-brane solution. In particular, the solution 
still has a regular horizon\foot{
This is to be contrasted with 
added pp-wave like terms $\Psi(u,r) \sim {1 \over r}$, or 
having plane waves along the longitudinal directions of the 
brane, wherein one does encounter singularities at the horizon {\it cf.},
\kmr, \gy , \bcr } at $r =0$. 
Furthermore, its asymptotic behaviour is that 
of a ten dimensional vacuum plane wave times an extra real line parameterized 
by $x$. 

In passing, we note that when $k \neq 0$, the nature of the asymptotics 
changes quite dramatically. The asymptotic behaviour of the 
equation \psimtworad\ can be analyzed to show that the solutions are 
Bessel functions, which are incompatible with the necessary $r^2$ behaviour.
So by looking for solutions wherein we have some momentum along the brane 
world-volume directions we do not obtain a solution that looks like a 
plane-wave.  In order to obtain a plane wave solution, our 
only choice is then to use the solution presented in \mtwodefsol, wherein 
we have an additional isometry corresponding to translations along 
$\( \partial \over \partial x\)^a$. We note that extensions of 
the GV technique were also considered in \lpv, \ops, and \bdgo.

Although the above solution \mtwodefsol, where we could compactify 
along the $x$ direction, is interesting in its own right,
in the bigger scheme of attempting to extend the BMN correspondence
to BFHP black hole spacetimes, it falls short in two important aspects:
it corresponds to an extremal, rather than a neutral, black hole;
and the asymptotic region corresponds to a vacuum, rather than the
maximally symmetric BFHP, plane wave.
In the next section we will rectify the first shortcoming completely
and the second one at least partially; but before presenting that
construction, we will make a few more remarks.

As a first step to correcting the above-mentioned deficiencies, 
one can try to look for a neutral black hole solution in 
asymptotically vacuum plane wave spacetime.
One would expect that there be such a solution, since after all, 
one might achieve it physically by colliding two oppositely charged
extremal black strings discussed above.
Indeed, assuming the same symmetries as for the ``seed''
asymptotically flat extremal black string (namely globally null 
Killing vector and transverse spherical symmetry), all vacuum solutions
can be found analytically.  This was done in \NullSoln, where the
new vacuum solutions were written down explicitly and their properties
analysed.  
Interestingly, it turns out that (independently of their asymptotic 
behaviour) none of these solutions can admit 
horizons; they are in fact nakedly singular.  
This exemplifies the remarkable point that restricting ourselves to 
vacuum Einstein's equations severely restricts the nature of the
possible solutions, and in particular eliminates the causally
nontrivial ones.

At the first sight, this result may seem paradoxical in light of the naive
argument of constructing vacuum black strings by colliding
oppositely-charged ones, but actually there is no contradiction:
the collision process would spoil the null symmetry.  In fact, this
might be expected already from the explicit construction outlined
above: only extremal black branes admit a null symmetry, so we can't 
start with a non-extremal solution instead.

The second shortcoming, that the asymptotics correspond to vacuum, rather
than the maximally symmetric plane wave (in the transverse directions), 
is in fact a necessary consequence of using the GV construction to generate
the solution
 from asymptotically flat spacetime: the matter content is unchanged, 
so vacuum asymptotics can only lead to vacuum asymptotics.  
In the next section, this will be contrasted by a different solution
generating technique, which in fact generates the maximally symmetric
(nonvacuum) plane wave starting from the flat space.  

\newsec{Plane wave black strings}

In the previous section we have seen that there exist asymptotically
vacuum plane wave solutions with horizons and null isometry; however, 
the black branes were extremal and the asymptopia non-maximally 
symmetric.  To do better, we should further relax the required symmetries
of the solution; namely, we now drop the globally null isometry.
As is often the case, the price we pay for breaking symmetries is 
manifested in the increased difficulty of finding the solutions.
Solving the ansatz by brute-force is forbiddingly messy; but fortunately,
one can once again resort to an elegant trick of using a particular
solution generating technique, which we call the ``Null Melvin Twist''
\PWBS.

Our method is based on the observation of \alga\
that certain class of plane wave geometries can be generated by
applying a sequence of manipulations to Minkowski space.  
By applying the same sequence of
manipulations starting from a black string solution,
we are able to generate a large class of black string deformations of
plane wave geometries with a regular horizon.  These solutions are
characterized by the mass density of the black string and the scale of
the plane wave geometry in the rest frame of the black string.  For
dimensions greater than six, the deformation due to the presence of
the black string decays at large distances in the direction transverse
to the string. Hence in these cases, our solutions describe geometries
with (radially) plane wave asymptotics. However, the solutions described 
will be plane-wave geometries which are supported by 3-form flux 
rather than 5-form flux as is desirable for the BMN correspondence. 

The specific set of steps is quite simple, and requires only a starting
10 dimensional supergravity solution (in NS-NS sector) with a
constant-norm isometry $\dya$.  One then performs the following sequence
of steps:

\itm
Boost the space-time along $y$ by $\g$

\itm
T-dualize along the $y$ coordinate

\itm
Twist by making a change of coordinates\foot{
 Here we write the
 metric on the  $\S^7$ in terms of polar coordinates
$ d\Om_7^2 = d\chi^2 + {1 \over 4} \, \cos^2 \chi \, d\Om_{a}^2
+ {1 \over 4} \,  \sin^2 \chi \,  d\Om_{b}^2$ 
with $ d\Om_{i}^2 \equiv d\theta_i^2 + d \psi_i^2 +  
d\phi_i^2 + 2 \cos \theta_i \, d\psi_i \, d\phi_i$, $i = a, b$,
and introduce a symmetric 1-form
$\sigma \equiv \cos^2 \chi \, 
( \cos \theta_1 \, d\psi_1 + d \phi_1 )
 + \sin^2 \chi \, ( \cos \theta_2 \, d\psi_2 + d \phi_2 )$.}
$\sigma \to \sigma + 2 \, \a \, dy$ 

\itm
T-dualize back along $y$

\itm
Boost back along $y$ by $-\g$

\itm
Finally, perform a double scaling limit, wherein the boost
$\g$ is scaled to infinity and the twist $\a$ to zero keeping
$ \b \equiv {1 \over 2} \, \a \, e^\g = $ fixed.

As mentioned above, starting with the 10 dimensional Minkowski spacetime,
$ 
ds_{str}^2  = - dt^2 + dy^2 + dr^2 + r^2 \, d\Om_7^2 $  
with $\Phi = 0, \ B = 0$,
we obtain the plane wave
\eqn\planety{\eqalign{
ds_{str}^2  &= - (1 + \b^2 \, r^2) \, dt^2 - 2 \, \b^2 \, r^2 \, dt \, dr
 + (1 - \b^2 \, r^2) \, dy^2  + dr^2 + r^2 \, d\Om_7^2 \cr
\Phi &= 0, \qquad B = {\b \, r^2 \over 2} \, \( dt + dy \) \wedge \sigma
}}
where the usual form is obtained by letting
$u = t + y$ and $2v = t - y$.

Similarly, starting with the asymptotically flat
 Schwarzschild black string solution,
\eqn\Schwbs{
ds_{str}^2  =  -f(r) \, dt^2 + dy^2 + {1 \over f(r)} \, dr^2 + r^2 \, d
\Om_7^2 }
one obtains 
\eqn\tendbstr{\eqalign{
ds_{str}^2 & = - {f(r)\, \(1 + \b^2 \, r^2
\) \over k(r)} \, dt^2 -  \, {2 \, \b^2 \, r^2 \, f(r) \over k(r)} \,
dt \, dy + \( 1  -{ \b^2\,  r^2 \over k(r)} \) \, dy^2  \cr
&  \qquad  + {dr^2 \over f(r)} + r^2 \, d\Om_7^2 - {\b^2 \, r^4 
\, (1 - f(r)) \over 4 \, k(r)}\, \sigma^2  \cr
e^{\Phi} &= { 1 \over \sqrt{k(r)}}\ , \qquad
B  = {\b r^2 \over 2 k(r)}\, \(f(r) \, dt + dy\) \wedge \sigma }}
where $f(r) \equiv 1 - {M \over r^6}$
and $k(r) \equiv 1 + {\b^2 M  \over r^4} $.

The solution \tendbstr\ is very simple. By inspection, if we
set $M$ to zero the solution reduces to the maximally symmetric
plane wave $\CP_{10}$.  
On the other hand,
setting $\b=0$ will reduce the solution to the black string
solution \Schwbs.  
For all finite values of $M$, there is a curvature singularity at 
$r=0$, which can be easily checked to be spacelike. 
More importantly, there is a regular horizon at
\eqn\horrad{
 r_H = M^{1/6} }
which persists for finite values of $\b$.  One can therefore
interpret \tendbstr\ as the black string deformation of
$\CP_{10}$.  Furthermore, since both $f(r)$ and $k(r)$ asymptote to 1
as $r$ is taken to be large, the effect of $M$ decays at large
$r$. Unlike the six dimensional solution described in \giha\
which deformed the geometry by a finite amount at large $r$,
\tendbstr\ is a black string solution which genuinely asymptotes
to $\CP_{10}$ (in the transverse directions). 

A rather remarkable property of this solution concerns its thermodynamics.
As discussed in \PWBS, the area of the horizon in Einstein frame metric
is given by
\eqn\area{
{\cal A}_H = L \, M^{{7/6}} \, \Om_7 \ ,}
while a natural definition\foot{
As discussed in \PWBS, there is a normalization ambiguity.}
of temperature would evaluate to 
\eqn\temp{
T_H = {3 \over 2 \pi} M^{-1/6} \ . }
The first law of thermodynamics would then naturally allow us to
interpret the parameter $M$ as the mass density of the black string.
The remarkable fact about these quantities is that they are  
independent of the parameter $\b$, \ie the plane wave black string
thermodynamics is the same as the usual asymptotically flat black
string thermodynamics.

As discussed in \PWBS, 
the solution \tendbstr\ can be generalized to include rotations, charge,
or more general twists, giving rise to a 13-parameter family of solutions,
all with regular horizons and maximally symmetric plane wave asymptotics.
Furthermore, one can also repeat the construction starting with black 
D$p$-branes, which generates plane wave $\CP_{11-p}$ asymptotics transverse
to the brane.  

Although these are all intriguing generalizations with
diverse and interesting properties, for the purposes of extending the
BMN story, it would be far more useful to generate solutions with
asymptotics which are not only metrically $\CP_{10}$, but also have the
requisite matter content, namely the RR 5-form field strength. In our 
construction, we explicitly exploited the usual duality between 
off-diagonal metric components and magnetic fields, which is why we obtained 
solutions supported by NS-NS flux. (In IIB it is trivial to S-dualise to 
generate solutions supported by a combination of NS-NS and RR 3-form fluxes).
However, since there is no way to geometrically engineer RR 5-form fluxes, 
we will have to work harder to generate solutions supported by the same. 

Nevertheless, 
it is tempting to conjecture that the final solutions will have similar 
thermodynamic properties as the solutions described above. 
This observation 
is simply based on the fact that the deformation of asymptotics from 
flat space to plane wave form seems to have no apparent effect on the 
thermal properties.

\newsec{Discussion}

We have summarized a particular development of ideas, to date still
incomplete, to construct asymptotically BMN black hole spacetimes.
Having shown that pp-wave (or more specifically, plane wave) 
metrics cannot represent black holes, we have reviewed the construction
of asymptotically plane wave black holes.  The first solution-generating
method (the GV construction) we used relied heavily on null isometry
and preserved the matter content of the seed solution.  As such, it only
allowed us to construct extremal black strings with vacuum plane wave
asymptopia.  While interesting in its own right, a far more fruitful 
technique proved to be the Null Melvin Twist, 
which allowed us to generate a wide class of black strings with the metric 
asymptoting to the desired maximally symmetric plane wave. However, 
even in the latter construction, we only managed to generate solutions that 
were supported by 3-form fluxes (in ten dimensional supergravity). 

There are many open questions that remain to be discussed in this 
general story. One obvious extension is to obtain solutions that have 
5-form flux support. Another would be to find explicit black hole 
solutions rather than the black string solutions mentioned above.
These appear to be sufficiently complicated to be tackled by brute-force 
solution of Einstein equations and we are not aware of any appropriate
solution generating technique. 
If such a solution generating technique were to be uncovered, its utility 
in constructing general warped flux compactifications would be enormous. 

At the same time there are many interesting avenues that are amenable to 
further study taking off from the solutions presented here. 
For instance, it is very intriguing that the plane wave deformation of the
black string geometries appears to be ``irrelevant'' and it would be 
worthwhile to investigate
the generality of this result. One should also study the causal properties 
of black strings in plane wave geometries to figure out whether 
there are any remnants of the exotic features seen in the causal structures of 
plane waves.  Furthermore, these solutions provide an excellent setting to 
understand the notions of ADM mass and other conserved charges in backgrounds
which are asymptotically plane wave.


\vskip 1cm

\centerline{\bf Acknowledgements}
It is a great pleasure to thank Aki Hashimoto, Oleg Lunin and Eric Gimon 
for collaboration.
We gratefully acknowledge the hospitality of
the Aspen Center for Physics and the Kavli Institute for Theoretical Physics, 
where parts of this project were completed.
VH is supported by NSF Grant PHY-9870115, while MR acknowledges support
from the Berkeley Center for Theoretical Physics and also partial support
from the DOE grant DE-AC03-76SF00098 and the NSF grant PHY-0098840.

%
%

\listrefs
\end